\documentclass[12pt,preprint]{aastex}

\shortauthors{Okamoto et al.}
\shorttitle{Arising cool column by helical flux emergence}
\begin{document}

\title{A rising cool column as signature for helical flux emergence and formation of prominence and coronal cavity}
\author{\textsc{
Takenori J. Okamoto,$^{a}$$^{b}$\footnote{T. J. O. is supported by the Research Fellowships from the Japan Society for the Promotion of Science for Young Scientists.}
Saku Tsuneta,$^{a}$
Thomas E. Berger$^{b}$
}}
\affil{
$^{a}$National Astronomical Observatory, Mitaka, Tokyo, 181-8588, Japan\\
$^{b}$Lockheed Martin Solar and Astrophysics Laboratory, B/252, 3251 Hanover St., Palo Alto, CA 94304, USA
}
\email{joten.okamoto@nao.ac.jp}

\begin{abstract}

Continuous observations were performed of a quiescent prominence with the Solar Optical Telescope (SOT) on board the \emph{Hinode} satellite on 2006 December 23--24.
A peculiar slowly-rising column of $\sim10^{4}$ K plasma develops from the lower atmosphere during the observations.
The apparent ascent speed of the column is 2 km s$^{-1}$, while the fine structures of the column exhibit much faster motion of up to 20 km s$^{-1}$.
The column eventually becomes a faint low-lying prominence.
Associated with the appearance of the column, an overlying coronal cavity seen in the X-ray and EUV moves upward at $\sim$5 km s$^{-1}$.
We discuss the relationship between these episodes, and suggest that they are due to the emergence of a helical flux rope that undergoes reconnection with lower coronal fields, 
possibly carrying material into the coronal cavity.
Under the assumption of the emerging flux scenario, the lower velocity of 2 km s$^{-1}$ and the higher one of 20 km s$^{-1}$ in the column are attributed to 
the rising motion of the emerging flux and 
to the outflow driven by magnetic reconnection between the emerging flux and the pre-existing coronal field, respectively.
The present paper gives a coherent explanation of the enigmatic phenomenon of the rising column with the emergence of the helical rope, and its effect on the corona.
We discuss the implications that the emergence of such a helical rope has on the dynamo process in the convection zone.

\end{abstract}


\section{Introduction}

The magnetic configuration in and around a prominence is apparently very complicated.
A prominence consists of relatively cool chromospheric plasma at coronal heights. 
Prominences are frequently associated with larger coronal structures known as ``coronal cavities''
that appear as dark regions surrounded by brighter coronal emission.
A coronal cavity is in turn surrounded by a helmet streamer that forms a cusp above the coronal cavity.
These structures are often seen in white-light eclipse images and EUV observations of quiescent prominences (e.g., Engvold 1989).

Many researchers have investigated the magnetic configuration of prominences
(e.g., Priest et al. 1989; Hood \& Anzer 1990; Aulanier \& D\'emoulin 1998; van Ballegooijen 2004; Low \& Petrie 2005),
and it is suggested that a twisted magnetic flux tube supports the prominence material below the dark coronal cavities
(e.g., van Ballegooijen \& Martens 1989; Hudson et al. 1999; Amari et al. 2003; Zhang \& Low 2005; Fan \& Gibson 2006; Mackay \& van Ballegooijen 2006; Magara 2007).
The plasma density in coronal cavities may be lower than the ambient corona due to the magnetic pressure of the flux tube.
In fact, observations consistent with this supposition show that dark cavities have 30--40\% lower densities compared with the surrounding regions (Fuller et al. 2008; V\'asquez et al. 2009).

There are numerous observational studies on prominences (Martin 1998, and see the references therein)
with multi-wavelength (e.g., Kucera et al. 2003; Schmieder et al. 2004; Heinzel et al. 2008) and with high-spatial resolution (e.g., Engvold 1976; Lin et al. 2005; Berger et al. 2008; Chae et al. 2008; Berger et al. 2010),
and also on the relationship between cavities and coronal mass ejections (e.g., Yurchyshyn 2002; Mari\v{c}i\v{c} et al. 2004; Vr\v{s}nak et al. 2004; Gibson et al. 2006).
While fewer analyses on activities inside dark cavities are found, recent observations reveal that cavities can have coherent velocity structures along the line of sight with speeds of 5--10 km s$^{-1}$ (Schmit et al. 2009).
The relationship between a prominence and the overlying cavity still remains unclear
although some theoretical approaches have been attempted (e.g., Low \& Hundhausen 1995; Low 2001; Gibson \& Fan 2006).

In this paper, we present analyses of a quiescent prominence with a slowly-rising motion from the lower atmosphere observed by
the Solar Optical Telescope (SOT: Tsuneta et al. 2008; Suematsu et al. 2008; Ichimoto et al. 2008; Shimizu et al. 2008) on board the \emph{Hinode} satellite (Kosugi et al. 2007).
The phenomenon is accompanied by activity in the overlying cavity as seen in the X-ray and EUV observations 
by the X-ray Telescope (XRT: Golub et al. 2007; Kano et al. 2008) aboard \emph{Hinode} and the Extreme Ultraviolet Imager (EUVI: W\"ulser et al. 2004) 
of the Sun Earth Connection Coronal and Heliospheric Investigation (SECCHI; Howard et al. 2008) aboard the NASA's Solar Terrestrial Relations Observatory (STEREO).
We investigate the episode in terms of the broader picture, namely the evolutionary process of the quiescent prominence and its relation to the overlying coronal cavity.
The observations suggest that an emerging helical flux rope is the fundamental driver of the episode.

\section{Observations}

The \emph{Hinode} satellite observed the north-west limb of the Sun from 11:23 UT on December 23 to 15:44 UT on December 24, 2006.
We obtained continuous images of a quiescent prominence with the Ca \textsc{ii}~H-line filter (3968\AA, bandwidth: 3\AA) of the SOT.
The field of view is 216\arcsec$\times$108\arcsec (2048$\times$1024 pixel$^2$) and the cadence is 10 seconds with brief interruptions for synoptic observations.
Coronal images of $\sim$2-MK plasma were taken with the XRT.
The Al/Polyimide filter combination was used with a cadence of between 19 and 70 s,  except for several observing gaps; the FOV is 512\arcsec$\times$512\arcsec.
We also use EUV datasets observed with the 195\AA\ and 304\AA\ filters of the STEREO/SECCHI/EUVI to identify coronal structures located over the prominence. 
The 195\AA\ filter images coronal emission primarily from the Fe~\textsc{xii} spectral line while the 304\AA\ filter images transition region plasma in the He~\textsc{ii} line.
The EUVI obtained full-disk Sun images with a 5-minute cadence.
We summed six images of the EUVI 195\AA\ and twenty images of the XRT/Al-Poly channels in order to see faint features within the dark cavity more clearly.
In this study, we deal with images taken by both STEREO-A/B spacecrafts with no distinction,
because the two spacecrafts had a heliocentric separation of 0.011$^{\circ}$ at the time of these observations.

\begin{figure}
\epsscale{1.0}
 \plotone{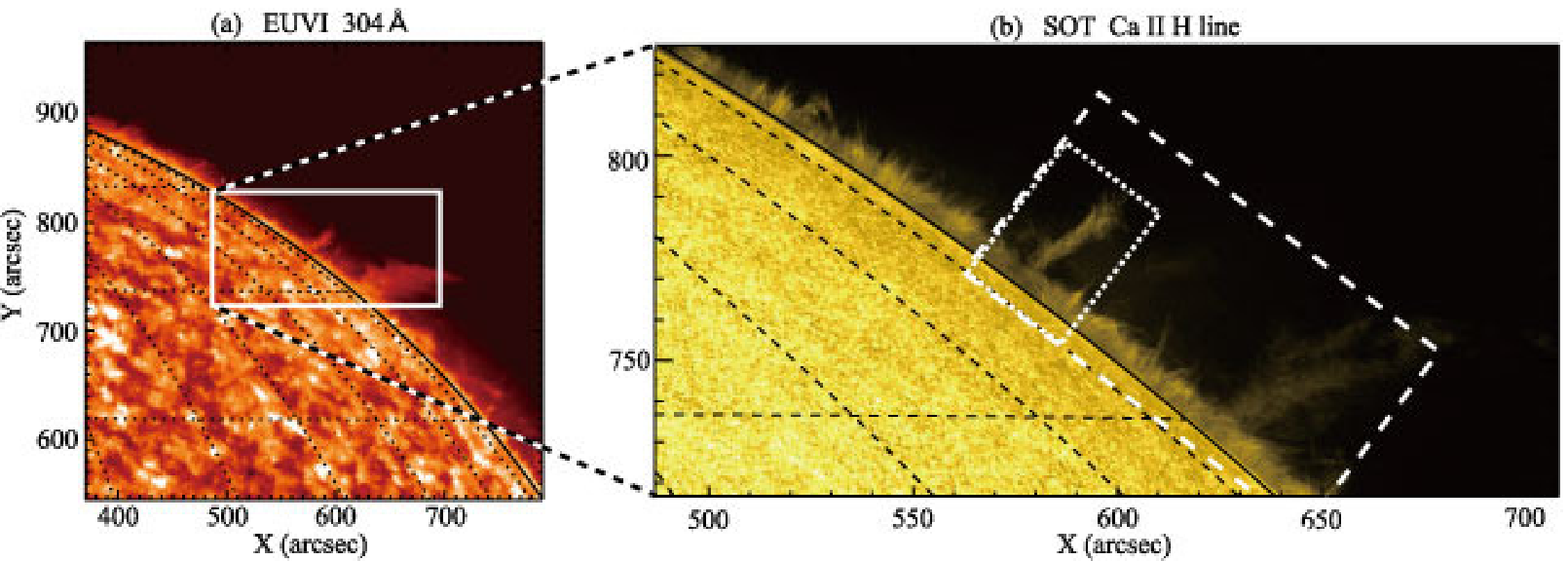}
    \caption{
Snapshots obtained at 13:25 UT on December 23, 2006. 
(a) 304\AA\ image observed with STEREO/SECCHI/EUVI.
(b) Ca \textsc{ii}~H line observed with \emph{Hinode}/SOT.
The location of the column is at (90W, 54N).
North is up, and East is to the left.
}
    \label{fig1}
\end{figure}

Figure \ref{fig1} shows EUVI 304\AA\ and SOT Ca \textsc{ii}~H-line snapshots observed at 13:25 UT on December 23.
There is a large prominence in the right side of the SOT image.
A time series of EUVI images shows that this was part of a polar crown prominence that erupted before the \emph{Hinode} observations started.
On the other hand, we can find a skinny vertical structure near the center of the FOV in the SOT.

\begin{figure}
\epsscale{1.0}
 \plotone{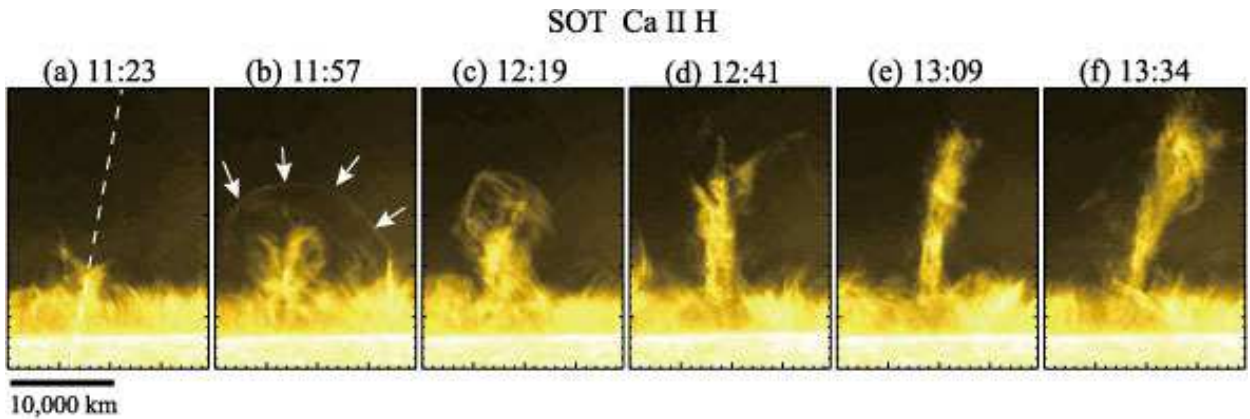}
    \caption{
A time-series of images obtained with the SOT.
The FOV is the same as the white-dotted box in Figure \ref{fig1}(b).
Tickmarks have a spacing of 1,000 km on the Sun.
A column-like structure extended into the corona with time.
The column consisted of numerous fine structures.
A circular-shaped structure was seen over the top of the column in the panel (b).
Apparent twisted motions were also observed in the column (i.e., in the panel (e)).
[This figure is available via two movies in the electronic edition of the Journal.]
}
    \label{fig2}
\end{figure}

Figure \ref{fig2} is a time-series of the Ca \textsc{ii}~H-line data (see also online Movie 1).
The FOV is indicated with the white-dotted box in Figure \ref{fig1}(b).
At the beginning of the \emph{Hinode} observations, there was a short feature over the limb (Fig. \ref{fig2}(a)).
The structure started to extend upward slowly, and evolved into an elongated vertical structure.
Faint loop-like structures appear several times over the column around 12:00 UT (Fig. \ref{fig2}(b) and see also online Movie 2).
The column exhibits fine structure motions with continuous and rapid upward motions.
Figure \ref{fig3} is a height-time plot along the slit indicated with the white-dashed line in Figure \ref{fig2}(a).
The top of the column rose with a mean speed of 2 km s$^{-1}$, while numerous small components of the column had much higher speeds.
Some of the faster motions reach speeds of 20 km s$^{-1}$, but they appear to be suppressed by an invisible boundary corresponding to the top of the column.
Some of the lofted material falls down from the top of the column, while most of material appears to stay in the corona.

The growth of the column persisted for approximately 2 hours.
For the following several hours, the material lofted into the corona by the column moved somewhat horizontally (Fig. \ref{fig4}: the white-dashed box corresponds to Fig. \ref{fig2}(f)), 
to become a very faint quiescent prominence.
There was a preexisting prominence nearby, but the horizontal column passed in front of or behind the prominence without interaction.
We can verify the situation in the EUVI 304\AA\ movie (Figure \ref{fig5}(a1-a5), and see also online Movie 3).
The rising column is indicated with the white arrow in Figure \ref{fig5}(a2),
and starts to move to the right without interacting with the preexisting prominence and a low-lying hot loop (Fig. \ref{fig5}(c2--c5)).
The dark feature in 195\AA\ indicated with the white arrow in Figure \ref{fig5}(b2) also corresponds to the column observed with the Ca \textsc{ii}~H line,
while the X-ray data shows no counterpart of the column or no remarkable activity around it.
This implies that the column itself has a temperature of $\sim10^4$ K (Schrijver et al. 1999; Anzer \& Heinzel 2005; Gilbert et al. 2006), 
and the material in the column ascends while maintaining a low temperature.

We observe some activity in the overlying coronal cavity after appearance of the column.
Before 11:30 UT, there is a dark cavity and a bright core in the corona (Fig. \ref{fig5}(b1), and see also online Movie 3).
After the appearance of the column, the dark cavity located over the column becomes darker and moves up (Fig. \ref{fig5}(b3)). 
The dark cavity appears to expand (Fig. \ref{fig5}(b4)) and move up further.
Finally, it appears to move along the outer edge of the coronal cavity (see online Movie 3).
Figure \ref{fig6} shows the detailed motions of the cavity in one hour time steps.
The darker cavity rises up and pushes the preexisting bright core with a speed of up to 5 km s$^{-1}$.
It then expands rightward and becomes less dark.

\section{Interpretations}

The SOT observations reveal a rising motion of cool material from the lower atmosphere.
This is a very interesting phenomenon in terms of the origin of the cool material of prominences.
Here we consider three possible origins of the material: condensation of hotter coronal plasma, siphon flows, and a helical flux emergence (Patsourakos \& Vial 2002).

\begin{figure}
\epsscale{1.0}
 \plotone{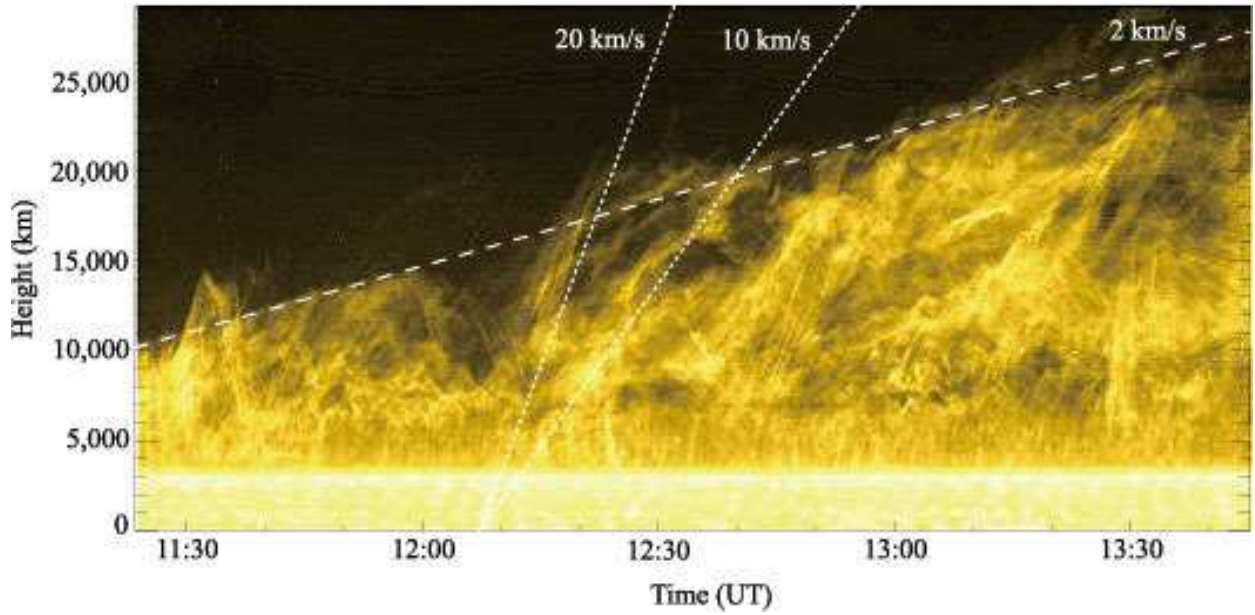}
    \caption{
Height-time plot along the slit indicated as the white-dashed line in Figure \ref{fig2}(a).
The horizontal axis is time (UT), and the vertical one is height along the slit.
The inclined lines delineate the rising structure, giving their speeds.
The top of the column rose with a mean speed of 2 km s$^{-1}$, while fine structures in the column had 5--20 km s$^{-1}$.
Some of them fell down after reaching the maximum heights, while the others went up with decelerating speed and stayed in the corona.
}
    \label{fig3}
\end{figure}

\begin{figure}
\epsscale{0.4}
 \plotone{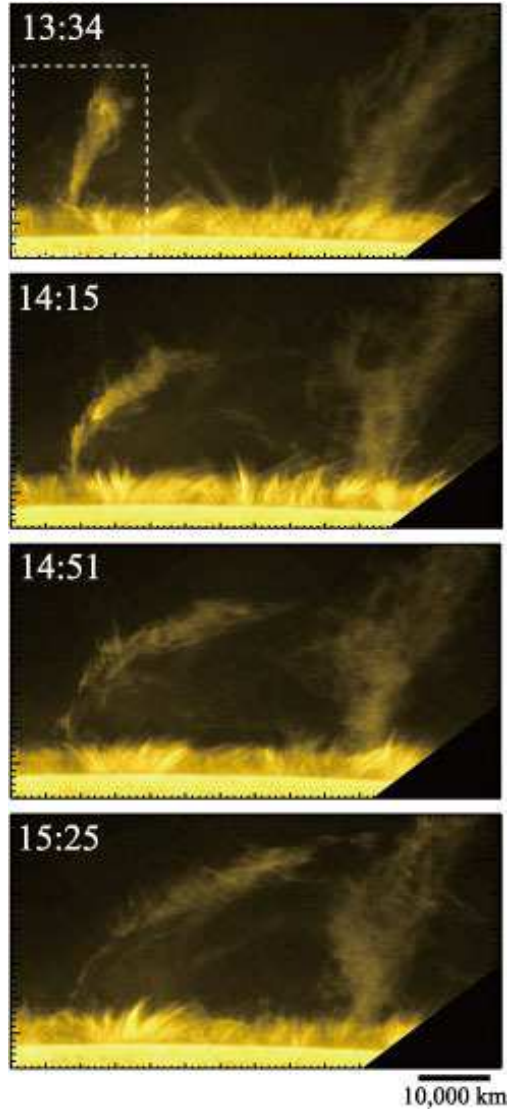}
    \caption{
A time-series of images obtained with the SOT.
The FOV is the same as the white-dashed box in Figure \ref{fig1}(b).
Tickmarks have a spacing of 1,000 km on the Sun.
The top panel includes the FOV of Figure \ref{fig2}(f) indicated as the white-dashed box.
The column moved horizontally, and passed in front of or behind the preexisting prominence without interaction.
}
    \label{fig4}
\end{figure}

\begin{figure}
\epsscale{0.8}
 \plotone{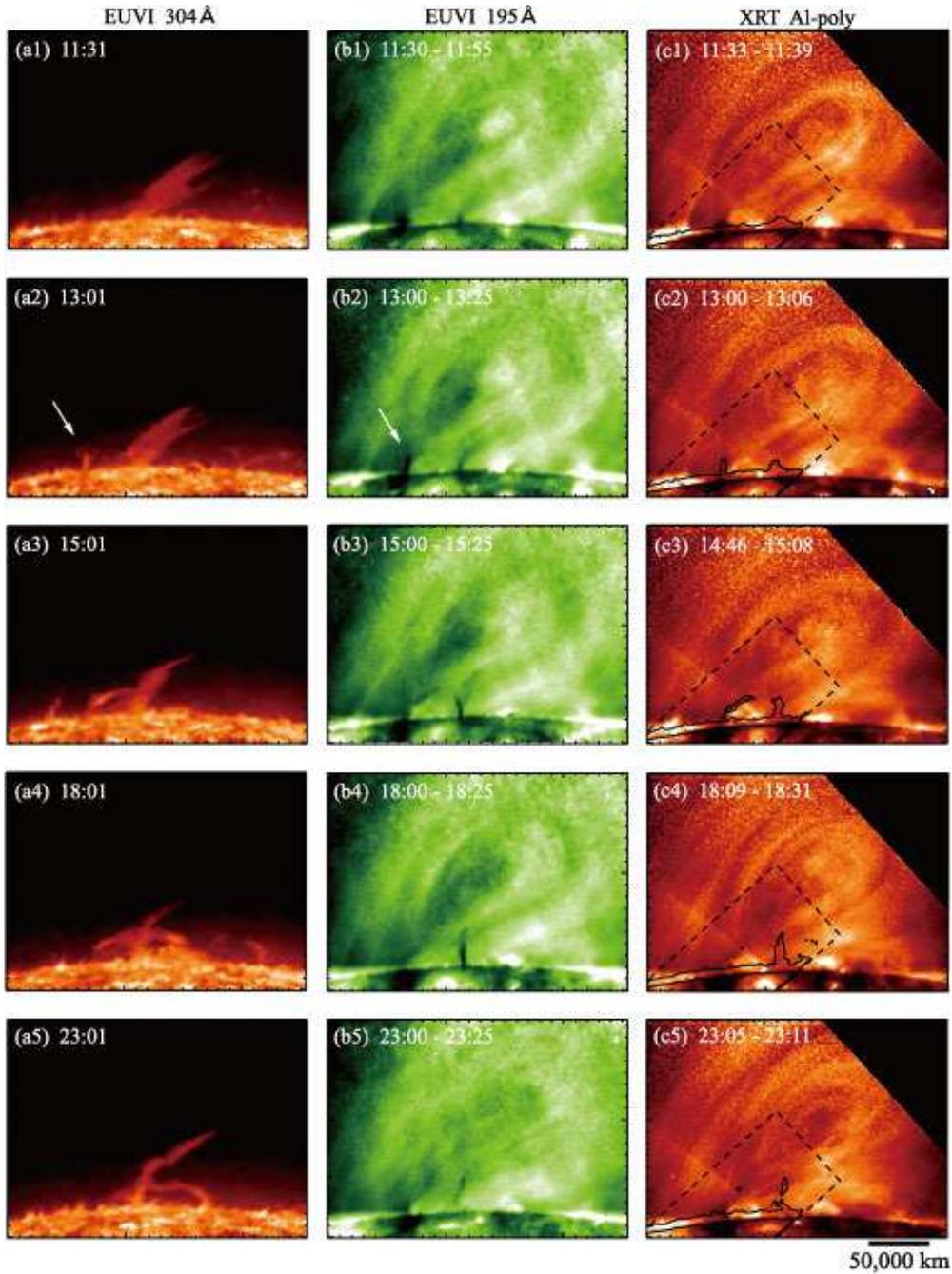}
    \caption{
A time-series of images in (left column) 304\AA, (center column) 195\AA, and (right column) X-ray.
Tickmarks have a spacing of 10,000 km on the Sun.
Each image in the panels (b1--b5, c1--c5) is integrated during a period shown in each panel.
The structures pointed by arrows in the panels (a2) and (b2) correspond to the column observed in the Ca \textsc{ii}~H line.
The dark cavity located over the column moved up after the appearance of the column, seen in the panels (b3--b5) and (c3--c5).
The black-dashed lines in the panels (c1--c5) are the FOVs of the SOT, and the black contours show the intensity of the Ca \textsc{ii}~H line.
The observed times of the SOT images are 11:33, 13:00, 14:46, 18:09, and 24:00 UT from above.
The coalignment between the SOT and the XRT was performed on the basis of Shimizu et al. (2007).
[This figure is available as a movie in the electronic edition of the Journal.]
}
    \label{fig5}
\end{figure}

\begin{figure}
\epsscale{1.0}
 \plotone{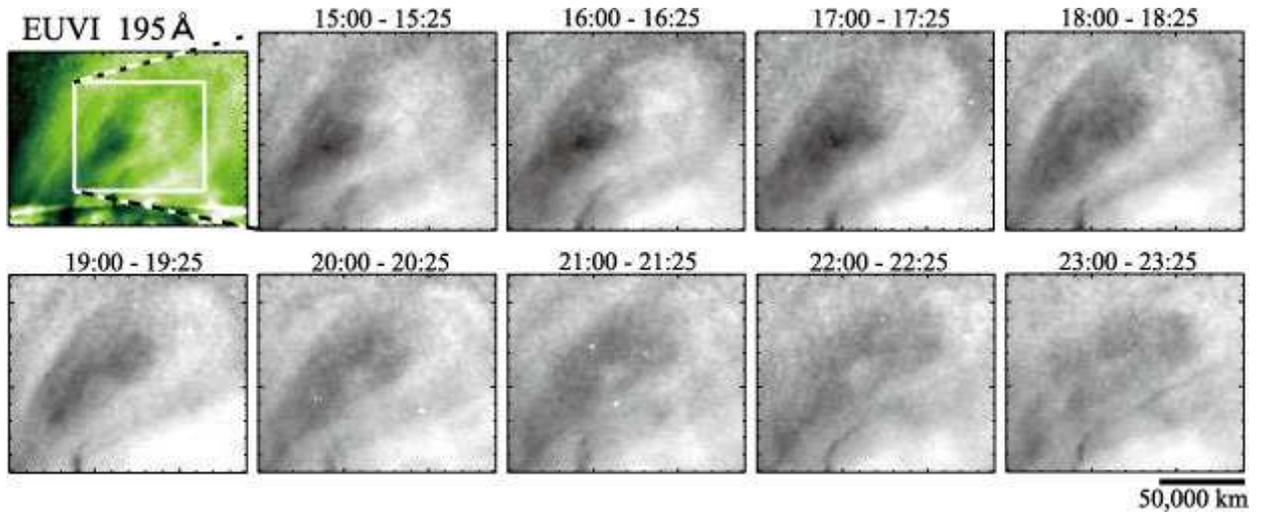}
    \caption{
A time-series of EUV images of the dark cavity.
Tickmarks have a spacing of 10,000 km on the Sun.
The top-left panel is the same as Figure \ref{fig5}(b3).
A dark region and a bright core are seen at the beginning.
The dark region moved up with 5 km s$^{-1}$ and pushed the bright core for 8 hours.
}
    \label{fig6}
\end{figure}

Kuperus \& Tandberg-Hanssen (1967) suggest that prominence mass comes from the overlying corona.
Thermal instability takes place in the current sheet located over the polarity inversion line on the photosphere,
and the condensed cool mass flows down to be trapped by coronal magnetic fields as an observed prominence (Malherbe et al. 1983; Pneuman 1983).
In our observations, however, we do not find cool material falling from above in the movie of the Ca \textsc{ii}~H line.
In contrast, we find cool material moving upward as shown in Figure \ref{fig3}.
Hence, coronal condensation does not seem to fit the observations.

Cool material may be brought up from the chromosphere with siphon flows (e.g., Pikel'ner 1971; Engvold \& Jensen 1977; Priest \& Smith 1979).
In this mechanism, a pressure difference between two footpoints drives a quasi-stationary siphon flow along the flux tube.
The material cools down as the flux tube expands and condenses to form a prominence.
This idea has been developed with numerical simulations (e.g., Dahlburg et al. 1998; Antiochos et al. 1999; DeVore \& Antiochos 2000; Karpen et al. 2001; Karpen \& Antiochos 2008).
In their studies, energy deposit to trigger the flow is localized at the footpoints of the magnetic fields in a prominence,
and the chromospheric plasma is heated up to several MK to evaporate into the corona.
Then, with thermal instability, the hotter plasma cools down, and becomes prominence material.
As long as the heating is intermittent, the prominence persists with the horizontal siphon flows seen as counterstreaming motions reported, e.g.,  by Zirker et al. (1998) and Martin (1998).
However, in our observations we do not see plasma with high temperatures at the footpoint or along the trace of the cool material in the EUV and X-ray observations (Fig. \ref{fig5}(b1--b3, c1--c3)): the plasma appears to be cool at all positions along the column. 
Since the material rises up as low-temperature $10^{4}$ K plasma, siphon flows are not be likely to be the key process in this phenomenon.

\begin{figure}
\epsscale{1.0}
 \plotone{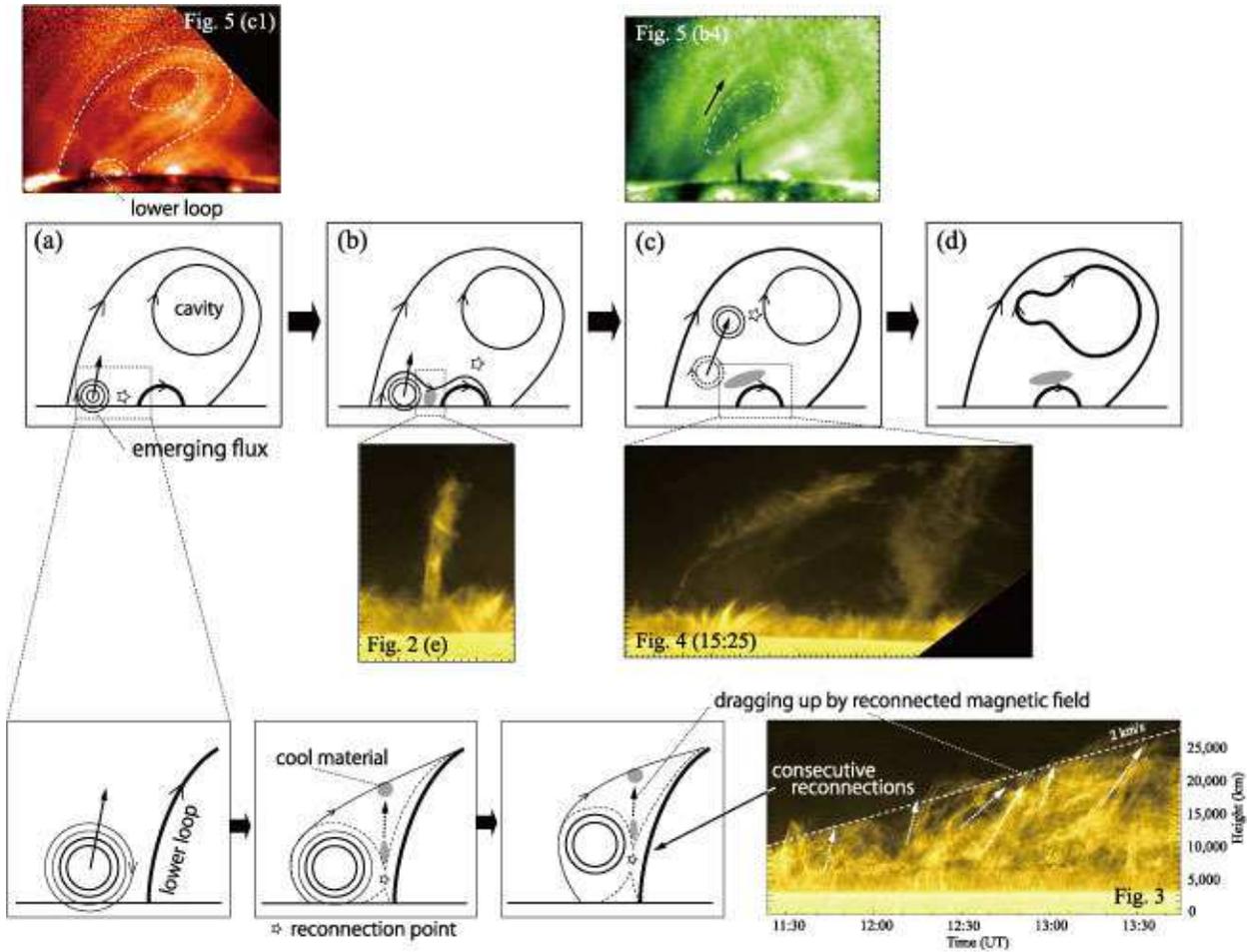}
    \caption{
Unified picture to explain all of the observations.
The star symbols indicate the locations of magnetic reconnection.
For simplicity, magnetic field lines are not shown here.
The detailed processes are depicted in the text.
Note that since we have no magnetograms, the directions of the magnetic fields are assumed.
}
    \label{fig7}
\end{figure}

Evidence for emerging helical flux ropes is seen in recent \emph{Hinode}/SOT observations (Okamoto et al. 2008, 2009; Lites 2009; Shimizu et al. 2009).
Unfortunately, we do not have any magnetograms in this study because the location is too close to the limb.
However, we believe that this mechanism best fits our observations and we thus present a unified interpretation that explains all of the key observational features. 
Figure \ref{fig7} depicts the proposed scenario for this episode.
First, we suppose that a helical flux rope emerges underneath the cavity (Fig. \ref{fig7}(a)).
If the direction of magnetic fields of the emerging flux rope is anti-parallel to that of the lower loop observed in the X-ray images (Fig. \ref{fig5}(c1)), reconnection will take place between them.
As a consequence, chromospheric materials are ejected upwards by the magnetic tension force of the reconnected field.
The maximum height of the lofted material depends on the altitude of the rising emerging flux, and corresponds to the top of the column (Fig. \ref{fig3}).
Consecutive reconnections at a rising X-point produce multiple cool material ejections, and the observed column is the envelop of such multiple ejections (Fig. \ref{fig7}(b)).
Here we have two types of velocities: 
The slower velocity of 2 km s$^{-1}$ is attributed to the rising motion of the emerging flux rope into the corona,
while the faster one of 20 km s$^{-1}$ is the ejection speed caused by magnetic reconnection between the emerging flux and the pre-existing coronal field.
The emerging flux remains coherent in spite of such repetitive reconnections, and expands in low coronal pressure region, reaching the pre-existing coronal cavity (Fig. \ref{fig7}(c)), consistent with the observations (Figs. \ref{fig5}(b4, c4) and \ref{fig6}).
Finally, the emerging flux system reconnects with the magnetic field of the pre-existing coronal cavity.
Thus we have a mechanism for carrying magnetic fields from the photosphere to maintain the cavity.

\section{Discussions}

\subsection{Implication of helical flux emergence}
As implied by Figure \ref{fig7}, the emergence of a flux rope system below a coronal cavity is consistent with our observations in Ca \textsc{ii}~H line, EUV, and X-ray. 
This scenario is based on both recent
\emph{Hinode} and ground-based observations that have shown several episodes consistent with emerging helical flux systems
(Lites et al. 1995; Okamoto et al. 2008; de Toma et al. 2008; Lites 2009; Shimizu et al. 2009; Berger et al. 2010).
Shimizu et al. (2009) reported the existence of the highly twisted magnetic flux tubes 
associated with intermittent but long-lasting chromospheric plasma ejections along a light bridge inside an umbra.
They suggested that the presence of helical flux tubes is essential for the observed phenomena,
creating antiparallel magnetic configurations favorable for magnetic reconnection to take place at one side of the emerging rope.
We point out that the chromospheric jets reported by Shimizu et al. (2009) are similar to 
the fine structures inside the upward motions of the column in our observations as shown in Figure \ref{fig3}.
In the case of Shimizu et al. (2009), the recurrent chromospheric ejections due to magnetic reconnection took place on one side of the light bridge 
at which magnetic fields are antiparallel between the umbra and vertical component of the helical flux.
Similarly, the column in our observations appears at one side of the possible emerging flux location.
The helical flux rope extends to the lower coronal loop during the emergence phase for several hours and intermittent reconnection between the flux rope and the loop drives jets, presumably supplying mass directly into the corona in this process.
In addition, we speculate that the circular and rotation-like motions seen in the Ca \textsc{ii}~H-line movie (online Movie 2 and Figure \ref{fig2}) 
are parts of the reconnected magnetic configurations consisting of the emerging helical flux and the pre-existing loop similar dynamics are seen by Lites \& Low (1997) and Inoue \& Kusano (2006).

Numerous observations of prominences have revealed the existence of barbs (e.g., Martin 1998).
Barbs are supposedly connected to patches of minority polarity on each side of a prominence (Martin \& Echols 1994).
Magara (2007) suggested that the formation of barbs is associated with emerging twisted flux,
and the helical flux rope reported in this paper could be the source of barbs and minority polarity patches.
Our dateset is, however, insufficient for discussion about the relationship between barbs and the emerging flux.
Further observations on the disk with the \emph{Hinode}/Spectro-Polarimeter will address the issue.

\subsection{Rising speed of the column}

Here we compare the peculiar slowly-rising motion observed in the Ca \textsc{ii}~H line to cases of prominence eruptions.
Numerous authors report that prominences rise slowly at about 0.1--1~km s$^{-1}$ several hours before eruptions 
(e.g., Sterling \& Moore 2004; Isobe \& Tripathi 2006; Nagashima et al. 2007; Isobe et al. 2007).
The apparent speed in our observations is similar to these values.
The episode shown in this paper is, however, different from these prominence eruptions simply
because no eruption of the column or the coronal cavity was observed in our case.
The size and apparent shapes of the column are also similar to those of the erupting prominence 
reported by Kurokawa et al. (1987) and one of the ``solar tornado'' events reported by Pike \& Mason (1998).
Nevertheless, the velocity of the column is several factors or one order of magnitude slower than in those cases.

Under the assumption of the emerging flux scenario, 
the slower velocity of 2 km s$^{-1}$ reflects the rising motion of the emerging flux in the corona.
The upward velocity of the coronal cavity is approximately 5 km s$^{-1}$, which we assume is driven by the rising column.
The measured speeds are also consistent with the results in numerical simulations of flux emergence (e.g., Yokoyama \& Shibata 1996).

\subsection{Fine structure speeds in the column}

Higher velocities of around 10 km s$^{-1}$ are observed in the fine structures of the column.
We conjecture that these are reconnection outflows as shown in Figure \ref{fig7}.
The velocities are consistent with the Alfv\'en velocity in the quiet Sun of 10--30 km s$^{-1}$,
where the magnetic field strength is an order of 10 G, and the electron number density is 10$^{11-12}$ cm$^{-3}$ as a value of quiescent prominence (Engvold 1990).
The upward reconnection outflow would have the Alfv\'en speed near the reconnection point, 
while the gravity force may decelerate the outflow speed.

We note that the observed height of the fine structures cannot be reached with the reconnection outflow impulse alone.
The heights of the fine structures were as high as 5,000--10,000 km, which is much higher than the maximum height of $\sim 800$~km achieved by  ballistic motion with an initial speed of 20~km~s$^{-1}$.
Hence, the materials must be pulled up by magnetic tension forces (Fig. \ref{fig7}) or ejected as a result of the propagation of slow-mode shocks (Shibata et al. 1982).


\subsection{Prominence mass supply}

In our observations, a new prominence was created as a result of slowly-rising motions of the cool materials from the lower atmosphere.
Berger et al. (2010) review the ``prominence mass problem'' in the context of their analysis of quiescent prominence plumes. They point out that the traditional mechanisms for supplying mass into prominences, namely coronal condensation and footpoint siphon flows, are apparently insufficient to counter the mass lost to gravitational drainage in quiescent prominences. Here we observe the initial formation phases of a quiescent prominence structure and propose that ``direct injection'' via magnetic reconnection may be a primary mechanism for supplying the initial mass into prominences. 

Direct injection has been observed in other contexts in the solar atmosphere. For example, Zirin (1976) and Liu et al. (2005) report observations of injections of cool material into the corona by surges, although the scale of surges is significantly larger than that of our episode. On a smaller scale, spicules have been suggested as a magnetically driven mass source for the solar wind (de Pontieu et al. 2009).
These authors all suggest that the frequent injection of mass through magnetic reconnection driven flows is an important transport mechanism in the outer solar atmosphere. 
Wang (1999) also suggests that quiescent prominences are organized systems of chromospheric jets triggered by magnetic reconnection between submerging flux and prominence barbs.
This is consistent with numerical studies showing that magnetic reconnection in the temperature minimum region can lead to upward mass flux and prominence formation (Litvinenko \& Martin 1999).
Flux rope emergence and reconnection may be yet another mechanism by which cool material is injected into the corona to form prominences.

\subsection{Coronal cavity dynamics}

We find an upward motion of the dark coronal cavity following the appearance of the cool column in the lower atmosphere.
The apparent velocity of the cavity motion is $\sim$5~km s$^{-1}$.
This is similar to the line-of-sight velocities measured in dark cavities by Schmit et al. (2009).
We are supposing that emerging flux triggers the motion seen higher up in the cavity as the flux rope rises into the corona. 
We note that there was no eruption associated with observed coronal cavity motions.
Hence, it may be that the repetition of flux emergence events supplies mass and magnetic flux into coronal cavities through dynamical coalescence without eruptions, as suggested in Figure \ref{fig7}. 
Berger et al.  (2010) mention that the dark buoyant cavities seen rising into pre-existing quiescent prominences may also be a source of mass and flux for the overlying coronal cavities. 
We might suppose that the flux emergence event seen here would have resulted in a prominence cavity (and subsequent plume formations) 
and had there been an overlying prominence for the flux rope to rise into. 
However since the prominence had already erupted prior to this emergence event, the flux rope in this case simply rises directly into the coronal cavity without interacting with an overlying prominence. 
If the origin of coronal cavities is indeed a series of emerging flux ropes as proposed here, the existing cavity and the emerging flux ropes should have the same helicity, giving favorable conditions for magnetic reconnection.

\subsection{Flux emergence at various scales and its implication for dynamo process}

Recent \emph{Hinode} observations have revealed flux emergences in different circumstances and on differed scales:
granular-scale emergence in plage regions (Ishikawa et al. 2008) and in the quiet Sun (Centeno et al. 2007), 
helical flux emergence in a light bridge of a sunspot (Shimizu et al. 2009),
emergence producing polar jets in the polar regions (Shimojo \& Tsuneta 2009),
and large-scale helical flux emergence in active regions (Okamoto et al. 2008, 2009; Lites 2009).
Ishikawa \& Tsuneta (2009) discuss the proposition that a local dynamo process due to the granular-scale convection may generate small-scale magnetic fields all over the Sun. 
Horizontal magnetic flux ranging in size from the granule-scale horizontal fields to the larger scale flux rope systems like the one reported here could presumably all have their origin in the convection zone to produce helical flux systems. These flux systems may all 
play an important role in modulating coronal activity on various size scales from 1,000--30,000~km and time scales of a few minutes to several days. Thus magnetic effects in the outer solar atmosphere such as microflares, jets and surges, and coronal cavity/prominence systems may have a common physical process as their origin, namely the small-scale convective dynamo operating in the outer convection zone.  Since the small-scale dynamo is presumably not solar cycle dependent, this may offer an explanation as to the existence of coronal cavities throughout the minimum of the cycle, even when sunspot active regions and large-scale magnetic field configurations in the photosphere are absent. 

\

The authors thank B. C. Low, S. E. Gibson, M. Kubo, D. J. Schmit, and A. C. Sterling for useful discussions.
We also appreciate the Hinode Operation Team.
\emph{Hinode} is a Japanese mission developed and launched by ISAS/JAXA, with NAOJ as domestic partner and NASA and STFC (UK) as international partners.
It is operated by these agencies in co-operation with ESA and NSC (Norway).
The SECCHI data are produced by an international consortium of the NRL, LMSAL, and NASA GSFC (USA), RAL and U. Bham (UK), MPS (Germany), CSL (Belgium), IOTA, and IAS (France).
This work was supported by KAKENHI (21$\cdot$8014) and carried out at the NAOJ Hinode Science Center, 
which was supported by the Grant-in-Aid for Creative Scientific Research ``The Basic Study of
Space Weather Prediction'' from MEXT, Japan (Head Investigator: K. Shibata), generous donation from the Sun Microsystems Inc., and NAOJ internal funding.


\end{document}